\documentstyle[preprint,aps]{revtex}

\tightenlines
\draft
\begin{document}
\date{13 July, 1998}
\preprint{RUB-TPII-13/98}
\title{Sudakov Form Factors with an Effective Theory of Particle
       Worldlines \\}
\author{N. G. Stefanis \thanks{Invited talk presented at the
        XI International Conference {\it Problems of Quantum Field
        Theory}, in memory of D. I. Blokhintsev, July 13-17, 1998,
        JINR, Dubna, Russia. To be published in the Proceedings.}
        }
\address{Institut f\"ur Theoretische Physik II,             \\
         Ruhr-Universit\"at Bochum,                         \\
         D-44780 Bochum, Germany                            \\
         Email: stefanis@hadron.tp2.ruhr-uni-bochum.de      \\
         }
\maketitle
\begin{abstract}
Sudakov-type form factors for quark vertex functions in QCD are
discussed using a worldline approach.
\end{abstract}


\newpage 
\input amssym.def
\input amssym.tex

\section{I\lowercase{ntroductory remarks}}
A very convenient context for investigating the low-energy regime of
gauge field theories is provided by particle worldlines \cite{Pol79}.
I begin by recalling the basics of this approach, discussed in more
detail in a series of publications [2-5]
on fermionic Green and vertex functions in the infrared (IR) domain
of QED and QCD. Then I shall focus on the electromagnetic quark vertex
function in QCD and present recent results for Sudakov-type form
factors. A main point will be to show how one can obtain such results
avoiding complicated diagrammatic techniques \cite{CS81}.

\section{B\lowercase{asics of the worldline approach}}
The worldline approach is simply a framework for trading field degrees
of freedom for those of particles, i.e., symbolically,
\begin{equation}
  \int_{}^{}[d\bar{\psi}][d\psi]\,
  {\rm e}^{S_{\rm field}[\bar{\psi},\psi]}
\longrightarrow
  \int_{}^{}[dx(\tau )][dp(\tau )]
  {\rm e}^{S_{\rm particle}[x(\tau ),p(\tau )]} \; .
\label{eq:fieldpartconv}
\end{equation}
The treatment of the bosonic sector is to be adapted to the particular
dynamical situation and is encoded in the expectation value of the
Wilson line operator.
Once a soft sector of the theory in question has been isolated, i.e., a
factorization scheme has been established, the infrared behavior
of Green and vertex functions is determined in terms of universal
anomalous dimensions \cite{KR87}, pertaining to the specific geometry
of the dominant (particle) worldlines. In this way, physical conditions,
e.g., interactions, off-mass-shellness, etc. translate into worldline
obstructions, like cusps, end-points, etc. \cite{Ste96}.

\section{T\lowercase{hree-point function}}
The fermion (quark) form factor within the worldline approach is defined
as the functional derivative of the generating functional:
\begin{equation}
  {\cal G}_{\mu}^{ij}(x,y;z)\,
=
  -\frac{\delta}{\delta J_{\mu}(z)}
   \frac{\delta ^{2}}{\delta\bar{\eta}_{i}(x)\delta\eta _{j}(y)}
   \ln Z[\bar{\eta},\eta,J_{\mu}]\Big\vert {}_{\bar\eta,\eta,J_{\mu}=0}
\; ,
\label{eq:vertexdef}
\end{equation}
where $Z[\bar{\eta},\eta,J_{\mu}]$ is the well-known expression for a
non-Abelian gauge field theory.
After conversion into particle variables, the form factor reads
\begin{eqnarray}
  {\cal G}_{\mu}^{ij}(x,y;z)\,
& = &
  \,\int_{0^{+}}^{\infty} dT
  \int_{}^{}[dx(\tau )] \,
  \int_{}^{}[dp(\tau )] {\cal P}
  \exp\left\{ -\int_{0}^{s} d\tau
      \left[ i p (\tau ) \cdot \gamma\,+\,m \right]
      \right\}
  \Gamma _{\mu}
\nonumber \\
& &
  \times {\cal P}
  \exp\left\{
             - \int_{s}^{T} d\tau
               \left[ i p (\tau ) \cdot \gamma\,+\,m \right]
      \right\}
  \exp\left[ i \int_{0}^{T} d\tau \, p(\tau )
             \cdot \dot{x}(\tau )\right]
\nonumber \\
& &
  \times \left\langle {\cal P}
  \exp\left[ i g\int_{0}^{T} d\tau \, \dot{x}(\tau ) \cdot
  {\cal A}\left(x(\tau )\right) \right]^{ij}\right\rangle _{A} \; ,
\label{eq:3pGf}
\end{eqnarray}
which involves two fermion lines coupled to a color-singlet current of
the form
$\bar{\psi}(z)\Gamma _\mu\psi (z)$, and where $[dx(\tau )]$ is to be
evaluated under the constraints:
$x(0)=x, x(T)=y, x(s)=z$. Here $<\,>_{A}$ denotes functional averaging
of the bracketed quantity in the non-Abelian gauge field sector, while
$\cal P$ stands for path ordering of the exponential in connection
with $\gamma$-matrices and/or non-Abelian vector potentials.
Our basic computational task is the evaluation of the expectation value
of a Wilson line operator over paths from an initial point $x$ to a
final point $y$ obliged to pass through the point of interaction $z$
with the external current which injects the large momentum $Q^{2}$.
Note that in our approach such lines may have a finite length,
corresponding to off-mass-shell particles. A perturbative expansion
yields
\begin{eqnarray}
  \left\langle {\cal P}
  {\rm e}^{i g \int_{0}^{T} d\tau\dot{x}\cdot{\cal A}}
  \right\rangle _{A} \!
&\!\!\! = \!\!\!& \!
  1 + (i g)^2 \Biggl\{ \int_{0}^{s}d\tau _{1}
      \int_{0}^{s} d\tau _{2}\theta (\tau _{2}- \tau _{1})
    + \int_{s}^{T} d\tau _{1}\int_{s}^{T} d\tau _{2}
      \theta (\tau _{2}-\tau _{1})
\nonumber \\
& &
    +
  \int_{0}^{s}\!d\tau _{1}\! \int_{s}^{T}\!d\tau _{2}\!\Biggr\}
  \dot{x}_\mu (\tau _{1})\dot{x}_\nu (\tau _{2})
  \left\langle{\cal A}_\mu \left(x(\tau _{1})\right)\!
  {\cal A}_\nu \left(x(\tau _{2})\right)\right\rangle _A
  + \ldots
\label{eq:pexp}
\end{eqnarray}
In the Feynman gauge and for a dimensionally regularized casting, the
gauge field correlator reads
$
  \left\langle{\cal A}_\mu (x)\,
  {\cal A}_\nu (x')\right\rangle ^{{\rm reg}}_A
\, = \,
  \delta _{\mu \nu}\, C_{F}\, \left(\mu ^{4-D}/4\pi ^{D/2}\right)
  \Gamma \left(D/2-1\right)\, |x-x'|^{2-D}
$.

\section{F\lowercase{actorization of a soft sector}}
The next major step is to isolate the soft contribution to Eq.
(\ref{eq:pexp}). As mentioned, this is done for a particular worldline
geometry. To this end, we adopt a straight-line path going from $x$ to
$z$ (for which $\dot{x}(\tau )=u_1$ with $0<\tau <s$) and a second
such path from $z$ to $y$ (for which $\dot{x}(\tau )=u_2$ with
$s<\tau <T$).
The no-recoil situation entailed by this restriction, except at the
cusp point $z$, suggests that the active gauge-field degrees of freedom
entering our computation are bound by an upper momentum scale which
serves to separate ``soft'' from ``hard'' physics in our factorization
scheme.
The singularities at the cusp and the endpoints of the path
will give rise to renormalization factors with corresponding anomalous
dimensions. The latter will determine the renormalization-group
evolution of the form factor and eventually produce Sudakov-type form
factors.

Refraining from entering into technical details (see \cite{GKKS97}), I
present only results. Choosing a frame for which $|u_{1}|=|u_{2}|=|u|$
and using the boson field correlator given above to reexpress
Eq. (\ref{eq:pexp}) solely in terms of particle degrees of freedom, we
obtain in the limit $D\rightarrow 4_{-}$
\begin{eqnarray}
  \left\langle {\cal P}
  {\rm e}^{i g \int_{0}^{T} d\tau\dot{x}\cdot{\cal A}}
  \right\rangle _{A} \!
&\!\!\! = \!\!\!& \!
   1 - \,\frac{g^{2}}{4\pi ^{2}}\, C_{F}\,
       \Biggl\{
         \frac{1}{D-4}\,
         \left[\varphi(w)-2\right]
         \left[ 1 + \frac{D-4}{2}
                    \ln\left(\pi {\rm e}^{2+\gamma _{E}}\right)
         \right]
\nonumber \\
& &
\mathop{\phantom{1\,}}
   +\, \ln (\mu|u|s)
       \left[F_{4}\left(\frac{s}{T-s},w\right) - 1\right]
\nonumber \\
& &
\mathop{\phantom{1\,}}
   +\, \ln (\mu|u|(T-s))
       \left[F_{4}\left(\frac{T-s}{s},w\right) - 1\right]
\nonumber \\
& &
\mathop{\phantom{1\,}}
   +\, \frac{1}{D-4}
       \Biggl[\, F_{D}\left(\frac{s}{T-s},w\right)
   +   F_{D}\left(\frac{T-s}{s},w\right) - \varphi (w)
       \Biggr]
       \Biggr\} \; ,
\label{eq:poexpD4}
\end{eqnarray}
where $\gamma _{E}$ is the Euler-Mascheroni constant and
\begin{equation}
  F_{4}(x,w)
=
    \frac{w}{\sqrt{1-w^{2}}} \arctan\frac{\sqrt{1-w^{2}}}{w}
  - \frac{w}{\sqrt{1-w^{2}}}
\arctan\frac{\sqrt{1-w^2}}{x+w} \; ,
\label{eq:F4}
\end{equation}
\begin{equation}
  \varphi (w)
\equiv
  F_{4}(x,w) + F_{4}\left(\frac{1}{x},w\right)
=
  \frac{w}{\sqrt{1-w^{2}}} \arctan\frac{\sqrt{1 - w^{2}}}{w} \; ,
\label{eq:phi(w)}
\end{equation}
with the relative velocity $w$ given by
$
 w\equiv u_{1} \cdot u_{2}/|u_{1}||u_{2}|
=
 p_{1} \cdot p_{2}/|p_{1}||p_{2}|
$.

\section{E\lowercase{ffective low-energy theory}}
We have isolated a sector of the full theory in which the matter
particles (fermions) are ``dressed'' to a scale that makes them appear
extremely heavy to the active, in this low-energy sector, gauge field
degrees of freedom. This ``heaviness'' prevents their derailment from
the straight-line propagation -- in close analogy to a Bloch-Nordsieck
situation \cite{KS89} -- and inhibits the creation of
fermion-antifermion pairs (this corresponds to integrating out of the
theory hard gluons). This is nothing but eikonal behavior \cite{CS81}.
Hence the factorization scale of the full theory becomes now the UV
scale of the high-energy domain of this ``soft'' sector and UV
divergences give rise to anomalous dimensions, much like in HQETh.
The advantage is clear: IR divergences (of the full theory) can now be
treated by usual renormalization group techniques. The anomalous
dimensions are then just the coefficients of the leading UV
divergences in dimensional regularization and are due to the
obstructions of the worldline: the cusp (interaction point with the
external electromagnetic current) and the endpoints:
\begin{equation}
  \Gamma _{{\rm cusp}}
=
  \frac{\alpha _{s}}{\pi}\, C_{F}
  \left[
        \frac{w}{\sqrt{w^{2}-1}}
        \tanh^{-1} \left(\frac{\sqrt{w^{2}-1}}{w}\right) - 1
  \right] \; , \quad\quad\quad
  \Gamma _{{\rm end}}
=
  - \frac{\alpha _{s}}{2\pi}\, C_{F} \; ,
\label{eq:cuspends}
\end{equation}
where we continued back to Minkowski space and set:
$
 \frac{w}{\sqrt{1-w^{2}}}\arctan\frac{\sqrt{1-w^{2}}}{w}\,
\rightarrow \,
\newline
\frac{w}{\sqrt{w^{2}-1}}\tanh ^{-1}\left(\frac{\sqrt{w^{2}-1}}{w}\right)
$.
Hence, the total anomalous dimension of the finite cusped worldline is
\begin{equation}
  \gamma (\alpha _{s},w)
=
      \Gamma _{{\rm cusp}}
  + 2 \Gamma _{{\rm end}}
=
  \frac{\alpha _{s}}{\pi}\, C_F
  \left[\frac{w}{\sqrt{w^{2}-1}}
  \tanh ^{-1}\left(\frac{\sqrt{w^{2}-1}}{w}\right) - 2
  \right] \; .
\label{eq:cuspend}
\end{equation}

\section{O\lowercase{ff-mass-shell fermions}}
Now we return to the three-point function (Eq. (\ref{eq:3pGf})), take
its Fourier transform, and removing the pole in $D-4$, we get the
renormalized expression:
\begin{equation}
  {\cal G}_{\mu}(p_{1},p_{2})
=
  {\cal G}_{\mu}^{(0)}(p_{1},p_{2})
  \left[
          1
        - \ln\left(\mu |u|/\lambda\right)\gamma (\alpha _{s},w)
        - f(\alpha _{s},w)
    \right]\,
    +\, {\cal O}(g^{4}) \; ,
\label{eq:vertexmom}
\end{equation}
where ${\cal G}_{\mu}^{(0)}(p_{1},p_{2})$ is the product of two free
propagators along the two branches of the cusped worldline. Above,
$\lambda = |m^{2}-p^{2}|/m$
is an off-mass-shellness scale serving as an IR regulator, and
$
  Q^{2}
=
  - \left(p_{1}-p_{2}\right)^{2}
=
  - 2p^{2} + 2p_{1}\cdot p_{2} \; ,
$
or, equivalently,
$
  p_{1}\cdot p_{2}/|p|^{2}
=
  w
=
  1 + Q^{2}/2|p|^{2} \; ,
$
where $p_{1}$ and $p_{2}$ are, respectively, the four-momenta of the
initial and final quark with mass $m$.
The last term in on the rhs of Eq.~(\ref{eq:vertexmom}) is given by
$
  f(\alpha _{s},w)
=
  \lim_{D \rightarrow 4_{-}}
  \left[
        2F_{D}(1,w)-\varphi (w)
  \right]\,
        \frac{\alpha _{s}/\pi}{D-4}
$
and is actually finite.
Now, the reference mass scale $\mu$ can range from a minimum value
$\mu _{{\rm min}}$ all the way up to $\mu _{{\rm max}}\sim |Q|$.
Below $\mu _{{\rm min}}$, we demand that the three-point function
describes free propagation along the two branches of the worldline.
We are thus led to the equation
\begin{equation}
  \ln \left(\mu _{{\rm min}}\, |u|/\lambda\right)
  \gamma (\alpha _{s},w) + f(\alpha _{s},w)
=
  0 \; ,
\label{eq:gammaef}
\end{equation}
which can be solved for $\mu _{{\rm min}}$, when
$Q^{2}/p^{2}\to\infty$. Then
\begin{equation}
  \gamma (\alpha _{s},w)
\simeq
  \frac{\alpha _{s}}{\pi}\, C_{F}\, \ln\frac{Q^{2}}{p^{2}}
\hspace{0.1in} , \hspace {0.2in}
  f(\alpha _{s},w)
\simeq
  \frac{\alpha _{s}}{4\pi}\, C_{F}\, \ln ^{2}\frac{Q^{2}}{p^{2}} \; ,
\label{eq:gammafsol}
\end{equation}
whereupon we determine (see also \cite{KR87})
\begin{equation}
  \mu _{{\rm min}}
=
  |m^{2}-p^{2}|/(Q^{2}p^{2})^{1/4} \; .
\label{eq:mu_min}
\end{equation}
A similar result holds also for ${\cal G}_{\mu}(p_{1},p_{2})$
on-mass-shell, the only difference being that the contribution to the
anomalous dimension due to the end-points is absent \cite{GKKS97}.

\section{R\lowercase{enormalization group evolution}}
To simplify the discussion, I consider the on-mass-shell case and
refer for the off-mass-shell case to \cite{GKKS97}. Defining the
evolution operator by
$
 {\cal D}
\equiv
   \mu \partial / \partial\mu
 + \beta (g) \partial / \partial g
 + \Gamma _{\rm cusp} (w,g)
$
we have
\begin{equation}
  {\cal D} F_{\rm soft} \left(w, \mu ^{2}/\lambda ^{2}\right)
=
  0 \; .
\label{eq:rge}
\end{equation}
In the limit $w\rightarrow \infty$ we find -- in accordance with
\cite{KR87}
\begin{equation}
  \Gamma _{{\rm cusp}}(w,g)
\simeq
  \ln \left(\frac{Q^{2}}{p^{2}}\right)\,
  \Gamma _{{\rm cusp}}(g)
  \hspace{0.1in}, \hspace{0.2in}
  \Gamma _{{\rm cusp}}(g)
=
  \frac{\alpha _{s}}{\pi}\, C_{F} + {\cal O}(g^{4}) \; .
\label{eq:gammacusp}
\end{equation}
Imposing $F_{\rm soft}(w,1)=1$, we obtain in LL approximation
($|p|=|p_{i}|,\,i=1,2$)
\begin{eqnarray}
  F_{\rm soft}\left(\frac{Q^{2}}{p^{2}},
                    \frac{\mu ^{2}}{\lambda ^{2}}
              \right)
& = &
  \exp\left[
            - \ln\left(\frac{Q^{2}}{p^{2}}\right)
              \int_{\lambda ^{2}}^{\mu ^{2}}
              \frac{dt}{2t}\,
              \Gamma _{{\rm cusp}}(g(t))
       \right]
\nonumber \\
& = &
  \exp\left[
           - \frac{C_{F}}{2\pi}\,
             \frac{4\pi}{\beta _{0}}\,
             \ln\left(\frac{Q^{2}}{p^{2}}\right)\,
             \ln\left(\frac{\ln\frac{\mu ^{2}}{\Lambda ^{2}}}
             {\ln\frac{\lambda ^{2}}{\Lambda ^{2}}}
                \right)
       \right] \; ,
\label{eq:sff}
\end{eqnarray}
where $\mu$ now denotes the separation point between the soft and the
hard sectors of the theory.
Then the total form factor can be factorized according to
\begin{equation}
  F\left( Q^{2}/\lambda ^{2}, \alpha _{s}(Q^{2}) \right)
=
  F_{\rm hard}\left( Q^{2}/\mu ^{2}, \alpha _{s}(Q^{2}) \right)
  \otimes
  F_{\rm soft}\left(
             Q^{2}/\mu ^{2}, \mu ^{2}/\lambda ^{2}, \alpha _{s}(Q^{2})
       \right) \; ,
\label{eq:fact}
\end{equation}
and taking into account that
$
 \frac{\partial}{\partial \ln Q^{2}} \ln F_{{\rm S}}
=
 - \int_{\lambda ^{2}}^{\mu ^{2}}\,\frac{dt}{2t}\,
   \Gamma _{{\rm cusp}}(g(t))
$, we find for
$Q^{2}/\lambda ^{2}\rightarrow \infty$
the final result for the form factor:
\begin{equation}
  F\left(Q^{2},\lambda ^{2}\right)
=
  \exp\left\{
            - \frac{C_{F}}{2\pi}\, \frac{4\pi}{\beta _{0}}\,
              \ln\left(\frac{Q^{2}}{\Lambda ^{2}}
                 \right)\,
              \ln\left[\frac{\ln \left(Q^{2}/\Lambda ^{2}\right)}
              {\ln\left(\lambda ^{2}/\Lambda ^{2}\right)}
                 \right]
            + \frac{C_{F}}{2\pi}\, \frac{4\pi}{\beta _{0}}\,
              \ln\left(\frac{Q^{2}}{\lambda ^{2}}\right)
      \right\} \; .
\label{eq:offmsff}
\end{equation}

\section{S\lowercase{ummary}}
I have shown that it is possible to derive useful information about the
quark electromagnetic vertex in QCD without relying upon complicated
Feynman diagram techniques, and pointed out the main features of the
worldline techniques employed. Applications to hadronic form factors
using the presented formalism just started being explored \cite{SSK98}.

\bigskip \medskip

\centerline{{\bf Acknowledgments}}
\bigskip
I would like to thank my long-term collaborators A. Karanikas and
C. Ktorides for their valuable contributions, and the organizers of
the conference for their generous hospitality and support.

\end{document}